\documentclass[final,5p,twocolumn]{elsarticle}

\usepackage{amsmath}
\usepackage{ulem}
\usepackage{tikz}
\usepackage{amssymb}
\usepackage{graphicx}
\usepackage{amsmath}
\usepackage{amsfonts}
\usepackage{subfigure}
\usepackage{tikz}
\usepackage{pgfplots}
\usepackage{array}
\usepackage{xcolor}
\usepackage{tikz}
\usetikzlibrary{chains,shapes.multipart}
\usetikzlibrary{shapes,calc}
\usetikzlibrary{automata,positioning}

\newcommand{\ie}{i.e.,\ }
\newcommand{\eg}{e.g.,\ }

\newtheorem{theorem}{Theorem}
\newtheorem{lemma}[theorem]{Lemma}
\newtheorem{proposition}[theorem]{Proposition}
\newtheorem{corollary}[theorem]{Corollary}
\newdefinition{rmk}{Remark}
\newproof{proof}{Proof}
\newproof{pot}{Proof of Theorem \ref{thm2}}

\makeatletter
\def\@author#1{\g@addto@macro\elsauthors{\normalsize%
    \def\baselinestretch{1}%
    \upshape\authorsep#1\unskip\textsuperscript{%
      \ifx\@fnmark\@empty\else\unskip\sep\@fnmark\let\sep=,\fi
      \ifx\@corref\@empty\else\unskip\sep\@corref\let\sep=,\fi
      }%
    \def\authorsep{\unskip,\space}%
    \global\let\@fnmark\@empty
    \global\let\@corref\@empty  
    \global\let\sep\@empty}%
    \@eadauthor={#1}
}
\makeatother

\begin{document}

\begin{frontmatter}



\title{Flexibility in an asymmetric system with prolonged service time at non-dedicated servers}

\author[label1]{Yanting Chen}
\address[label1]{Business School, University of Shanghai for Science and Technology, Shanghai, China 200093 \fnref{label1}}
\ead{yantingchen@usst.edu.cn}

\author{Jingui Xie\fnref{label2}}
\ead{xiej@ustc.edu.cn}
\address[label2]{School of Management, University of Science and Technology of China, Hefei, China 230026}

\author{Taozeng Zhu\corref{cor1}\fnref{label3}}
\ead{zhutozng@mail.ustc.edu.cn}
\cortext[cor1]{Corresponding author}
\address[label3]{NUS Business School, National University of Singapore, Singapore 119245}


\begin{abstract}
The prolonged service time at non-dedicated servers has been observed in~\cite{song2020capacity}. Motivated by such real problems, we propose a stylized model which characterizes the feature of the prolonged service time at non-dedicated servers in an asymmetric system. We study the independent system, the full flexibility system and the partial flexibility system when the occupation rate of the system, the degree of the prolonged service time and the degree of the asymmetry are allowed to change. We show that under certain circumstances, the partial flexibility scheme outperforms the full flexibility system and the independent system in such a model. Our results also provide instructions on how to introduce flexibility when the service time at non-dedicated servers is prolonged in an asymmetric system.
\end{abstract}


\end{frontmatter}


\section{Introduction}\label{sec:introduction}

Empirical studies indicate that the hidden (negative) consequences exist when the customer is served at the non-dedicated server. For instance, when the patients are assigned from a ward whose designated beds are fully occupied to an available bed in a unit designated for a different service, it has been shown in~\cite{song2020capacity} that this ``off-service placement'' is associated with a substantial increase in remaining hospital length of stay, \ie a prolonged service time. This arouses the question that whether we should introduce flexibility into the system and which level of flexibility is preferred.

Deciding on the right level of flexibility is a classic operations management problem: should different types of customers be processed or served with dedicated or flexible capacity? Substantial progress has been made in understanding the flexibility over the last $30$ years. One important insight is that the choice between dedication and flexibility is not an ``all-or-nothing" proposition. The most celebrated chaining structure (\cite{jordan1995benefits}) which is categorized as a partial flexibility scheme is proved to work surprisingly well in a symmetric system in the sense that the performance of such a system is almost as good as that of a full flexibility system which usually involves enormous implementation cost (\cite{bassamboo2012flexibility}).

The heterogeneity in service times makes it difficult to compare different levels of flexibility. The variability in service times would lead to deterioration of the system performance in a flexible system instead of improving it; it has been shown in~\cite{buzacott1996commonalities,smith1981resource} that the dedicated system works better when the service variability is large. Moreover, the asymmetric system, \ie the arriving rates are different, is also often encountered in practical situations. Unfortunately, the reviewing papers~\cite{chou2008process,wang2019review} show that majority of the existing flexibility literature focus on the symmetric systems in which all arriving rates and service rates are identical. The incorporation of the two practical features (prolonged service time at the non-dedicated servers, the asymmetric system) would distinguish our research from the existing literature. Moreover, these practical features call for new comparison of different levels of flexibility. 

In this paper, we rely on a stylized loss model with $2$ servers and $2$ types of customers. Although this is rarely the case in the real world, as~\cite{jordan1995benefits} argues, understanding simple systems can provide insights into very realistic scenarios. Moreover, to keep the first attempt of this problem tractable is also compelling. 

Our main contribution in this paper is to compare different levels of flexibility in a system incorporating the two practical features (prolonged service time at the non-dedicated servers, the asymmetric system), which has not been considered in previous studies. Our analysis reveals that in such a stochastic system, the partial flexibility could ever outperform the full flexibility. In particular, the partial flexibility is preferred when the asymmetry is obvious and the service times at the non-dedicated servers are prolonged with certain degree. Moreover, we show that when the system gets busier, working independently is the optimal choice. These findings indicate that the decision makers should be more cautious in deciding how to introduce flexibility into a service system while the system is asymmetric and the service time at the non-dedicated server is prolonged.

\section{Problem description and model formulation}\label{sec:model}
We consider two types of customers, \ie Type $1$ and Type $2$ customers. The dedicated servers for Type $1$ and Type $2$ customers are Server $1$ and Server $2$, respectively. The arrival processes for Type $1$ and Type $2$ customers are two independent Poisson processes with parameters $\lambda_1$ and $\lambda_2$, respectively. Without loss of generality, we assume that $\lambda_1 = \lambda$ and $\lambda_2 = k \lambda$ where $k \in [0,1]$. The service time of any customer at the dedicated server is exponential with parameter $\mu$. The service time of any customer at the non-dedicated server is exponential with parameter $\gamma \mu$ where $\gamma$ is \emph{the prolonged coefficient}. Due to the inefficiency of dealing with the non-dedicated customers, the prolonged coefficient $\gamma$ satisfies $\gamma \in [0,1]$ naturally. For example, if a Type $1$ customer is allowed to enter Server $2$ when Server $1$ is busy, then the service rate of the Type 1 customer at Server $2$ is $\gamma \mu$. There is no buffer space in front of the servers, the arriving customer who finds the dedicated server and the non-dedicated server (if redirecting is allowed) busy would be lost. The arrivals and services are mutually independent. 

\begin{figure}
\begin{center}
\subfigure[]{\begin{tikzpicture}[scale = 0.6]
		
		\draw (2.75,-0.75cm) circle [radius=0.75cm];
		\draw (2.75,1.5cm) circle [radius=0.75cm];
		\draw[->] (3.5,-0.75) -- +(20pt,0);
		\draw[->] (3.5,1.5) -- +(20pt,0);
		\draw[->] (-2,-0.75) -- +(100pt,0);
		\draw[->,red] (-0.3,-0.75) -- +(50pt,60pt);
		\draw[->] (-2,1.5) -- +(100pt,0);
		\draw[->,green] (-0.3,1.5) -- +(50pt,-60pt);
		\node at (2.75,-0.75cm) {$\mu$, {\color{green}$\gamma \mu$}};
		\node at (2.75,1.5cm) {$\mu$, {\color{red}$\gamma \mu$}};
		\node at (-1,-0.5cm) {$k\lambda$};
		\node at (-1,1.8cm) {$\lambda$};
			\node[align=center] at (-0.5cm,2.5cm) {Type $1$ customers};
		\node[align=center] at (-0.5cm,-1.5cm) {Type $2$ customers};
		\node[align=center] at (2.8cm,3cm) {Server $1$};
		\node[align=center] at (2.8cm,-2.1cm) {Server $2$};
		\end{tikzpicture}\label{fig:full_flexibility}} 
\subfigure[]{\begin{tikzpicture}[scale = 0.6]
		
		\draw (2.75,-0.75cm) circle [radius=0.75cm];
		\draw (2.75,1.5cm) circle [radius=0.75cm];

		\draw[->] (3.5,-0.75) -- +(20pt,0);
		\draw[->] (3.5,1.5) -- +(20pt,0);
		\draw[->] (-2,-0.75) -- +(100pt,0);
		\draw[->] (-2,1.5) -- +(100pt,0);
		\draw[->,green] (-0.3,1.5) -- +(50pt,-60pt);
		\node at (2.75,-0.75cm) {$\mu$, {\color{green}$\gamma \mu$}};
		\node at (2.75,1.5cm) {$\mu$};
		\node at (-1,-0.5cm) {$k \lambda$};
		\node at (-1,1.8cm) {$\lambda$};
			\node[align=center] at (-0.5cm,2.5cm) {Type $1$ customers};
		\node[align=center] at (-0.5cm,-1.5cm) {Type $2$ customers};
		\node[align=center] at (2.8cm,3cm) {Server $1$};
		\node[align=center] at (2.8cm,-2.1cm) {Server $2$};
		\end{tikzpicture}\label{fig:partial_flexibility}}
\end{center}
\caption{\subref{fig:full_flexibility} The full flexibility system.~\subref{fig:partial_flexibility} The partial flexibility system. \label{fig:flexible_systems}}
\end{figure}

There are $3$ possible flexibility designs for the system. In the~\emph{independent system}, the two service systems behave independently. In the~\emph{full flexibility system}, a Type $1$ customers who finds Server $1$ available would enter Server $1$. When a Type $1$ customer finds Server $1$ is busy upon arrival, this Type $1$ customer would immediately go to Server $2$, if Server $2$ is idle at this moment, this Type $1$ customer would receive service at Server $2$, otherwise, this Type $1$ customer would leave the system immediately. The behaviour of the Type $2$ customer is defined similarly. In the~\emph{partial flexibility system}, the Type $2$ customers are not allowed to redirect to Server $1$ under any circumstances. In other words, we forbid the busier server to help the lighter system. The illustration can be found in Figure~\ref{fig:flexible_systems}.

Without loss of generality, we rescale the system by assuming that the arrival rates of Type $1$ and Type $2$ customers are $\rho$ and $k \rho$ respectively, and the service rates at the dedicated server and the non-dedicated server are $1$ and $\gamma$ respectively, where $\rho = \frac{\lambda}{\mu}$. We formulate the full flexibility and the partial flexibility systems into continuous-time Markov chains $X(t)$ and $Y(t)$ respectively. The loss system (even when redirecting is allowed) is always stable, see~\citep{kleinrock1975queueing}. The possible states for Server $1$ and Server $2$ are $0,1$ and $2$, which means idle, occupied by a Type $1$ customer and occupied by a Type $2$ customer, respectively. The balance equations for $X(t)$ and $Y(t)$ can be found in Appendix~\ref{apx:balance_equations}. We denote the stationary probabilities of $X(t)$ by $\pi(i,j)$ where $i,j \in \{0,1,2\}$. Moreover, we denote the stationary probabilities of $Y(t)$ by $p(m,n)$ where $m \in \{0,1\}$ and $n \in \{0,1,2\}$. We use $T_{is}$, $T_{fs}$ and $T_{ps}$ to denote the throughputs of the independent system, the full flexibility system and the partial flexibility system respectively. Due to the PASTA property (Poisson arrivals see time averages), see~\cite{wolff1982poisson}, we have
\begin{align}
T_{is} =& \rho (1 - \frac{\rho}{\rho + 1}) + k \rho (1 - \frac{k \rho}{k \rho + 1}), \label{eq:tis}\\
T_{fs} =& (k + 1)\rho (1 - \pi(1,1) - \pi(1,2) - \pi(2,1) - \pi(2,2)), \label{eq:tfs}\\
T_{ps} =& \rho (1 - p(1,1) - p(1,2)) + \notag \\
        &k \rho(1 - p(1,1) - p(1,2) - p(0,1) - p(0,2)). \label{eq:tps}
\end{align}

We will compare the different flexibility systems first for the special scenarios, \eg all service times are identical and the system is symmetric. For the special scenarios, we extend the previous research results on our model. Moreover, the analysis of the special cases would assist the comparison of the different flexibility systems when the system parameters are more general.

\section{Identical service times}
In this section, we consider the identical service times, \ie $\gamma = 1$. We first obtain the stationary distribution of \emph{the full flexibility system with identical service times} in the next proposition.
\begin{proposition}{\label{pro:distribution_gammais1}}
When the service times are identical, \ie $\gamma = 1$, the stationary probabilities of the full flexibility system are
\begingroup
\allowdisplaybreaks
\begin{align*}
  \pi^{\gamma = 1}(0,0) &= \frac{2}{k^2\rho^2 + 2k\rho^2 + 2k\rho + \rho^2 + 2\rho + 2},\\
  \pi^{\gamma = 1}(1,0) &= \frac{\rho(k^2\rho^2 + 2k\rho^2 + 6k\rho + \rho^2 + 4\rho + 4)}{2k^2\rho^2 + 4k\rho^2 + 6k\rho + 2\rho^2 + 6\rho + 4} \pi^{\gamma = 1}(0,0)
 ,\\
  \pi^{\gamma = 1}(0,1) & = \frac{\rho^2(\rho k^2 + 2\rho k + \rho + 2)}{2 k^2\rho^2 + 4k\rho^2 + 6k\rho + 2\rho^2 + 6\rho + 4} \pi^{\gamma = 1}(0,0),\\
 \pi^{\gamma = 1}(0,2) & = \frac{k\rho(k^2\rho^2 + 2k\rho^2 + 4k\rho + \rho^2 + 6\rho + 4)}{2k^2\rho^2 + 4k\rho^2 + 6k\rho + 2\rho^2 + 6\rho + 4} \pi^{\gamma = 1}(0,0),\\
 \pi^{\gamma = 1}(2,0) &= \frac{k\rho^2(2k + \rho + 2k\rho + k^2\rho)}{2k^2\rho^2 + 4k\rho^2 + 6k\rho + 2\rho^2 + 6\rho + 4} \pi^{\gamma = 1}(0,0),\\
 \pi^{\gamma = 1}(1,1) &= \frac{\rho^2}{2} \pi^{\gamma = 1}(0,0),\\
 \pi^{\gamma = 1}(1,2) &= \frac{k\rho^2(\rho + k\rho + 4)}{2\rho + 2k\rho + 4} \pi^{\gamma = 1}(0,0),\\
 \pi^{\gamma = 1}(2,1) &= \frac{k\rho^3(k + 1)}{2\rho + 2k\rho + 4} \pi^{\gamma = 1}(0,0),\\
 \pi^{\gamma = 1}(2,2) &= \frac{k^2\rho^2}{2} \pi^{\gamma = 1}(0,0).
\end{align*}
\end{proposition}
\endgroup
The result can be readily verified thus we omit the proof. Next, we obtain the stationary distribution of \emph{the partial flexibility system with identical service times}.
\begin{proposition}{\label{pro:partial_gammais1}}
When the service times are identical, \ie $\gamma = 1$, the stationary probabilities of the partial flexibility system are
\begin{align*}
 p^{\gamma = 1}(0,0) = & \frac{2\rho + k\rho + 2}{(\rho + 1)(k^2\rho^2 + 2k\rho^2 + 3k\rho + \rho^2 + 2\rho + 2)},\\
 p^{\gamma = 1}(1,0) = & \frac{\rho(\rho + k\rho + 2)}{2\rho + k\rho + 2}  p^{\gamma = 1}(0,0), \\
 p^{\gamma = 1}(0,1) = & \frac{\rho^2(\rho + k\rho + 2)}{(\rho + 2)(2\rho + k\rho + 2)} p^{\gamma = 1}(0,0),\\
  p^{\gamma = 1}(0,2) =&\frac{k\rho(6\rho + 2k\rho + k\rho^2 + \rho^2 + 4)}{(\rho + 2)(2\rho + k\rho + 2)} p^{\gamma = 1}(0,0),\\
  p^{\gamma = 1}(1,1) = &\frac{\rho^2(\rho + 1)(\rho + k\rho + 2)}{(\rho + 2)(2\rho + k\rho + 2)} p^{\gamma = 1}(0,0),\\
  p^{\gamma = 1}(1,2) = &\frac{k\rho^2(5\rho + 2k\rho + k\rho^2 + \rho^2 + 4)}{(\rho + 2)(2\rho + k\rho + 2)} p^{\gamma = 1}(0,0).
\end{align*}
\end{proposition}
The result can be readily verified thus we omit the proof. With the assistance of the stationary probabilities of the full flexibility system and the partial flexibility system, we are now able to compare the throughputs of the $3$ systems when the service times are identical.

When the service times are identical, the throughputs of the $3$ systems can be obtained by substituting the corresponding stationary probabilities into Equations~\eqref{eq:tis}-\eqref{eq:tps},
\begin{align*}
T_{is}^{\gamma = 1} =& \frac{2 k \rho^2 + (k + 1)\rho}{(\rho + 1)(k \rho +1)},\\
T_{fs}^{\gamma = 1} =& \frac{2\rho(k + 1)(\rho + k\rho + 1)}{k^2\rho^2 + 2 k\rho^2 + 2k\rho + \rho^2 + 2\rho + 2}, \\
T_{ps}^{\gamma = 1} =& \frac{\rho(2k^2\rho^2 + k^2\rho + 4k\rho^2 + 7k\rho + 2k + 2\rho^2 + 4\rho + 2)}{(\rho + 1)(k^2\rho^2 + 2k\rho^2 + 3k\rho + \rho^2 + 2\rho + 2)}.
\end{align*}
We then obtain the relationship of the throughputs when the service times are identical.
\begin{proposition}{\label{pro:gamma1}}
When the service times are identical, \ie $\gamma = 1$, we have $T^{\gamma = 1}_{is} < T^{\gamma = 1}_{ps} < T^{\gamma = 1}_{fs}$.
\end{proposition}

We defer all the proofs to the appendix. Our result matches the main conclusion in the existing literature~\cite{smith1981resource} stating that when all the service times are identical, in a symmetric system, the full flexibility is always beneficial. Moreover, our result extends this general conception to the asymmetric system as well.

The opposite extreme case of the case discussed above is $\gamma = 0$, \ie the customer which is redirected to the non-dedicated server would lead to extremely long service time. By intuition, when a redirecting occurs, the corresponding non-dedicated server would be fully blocked. Therefore, it is obvious that the independent system would be optimal in this situation. In fact, we can conclude that
\begin{proposition}{\label{pro:gamma0}}
When $\gamma = 0$, we have $T^{\gamma = 0}_{fs} < T^{\gamma = 0}_{ps} \leq T^{\gamma = 0}_{is}$, the equality sign holds when $k = 0$.
\end{proposition}
This result is also necessary in the proof later on. Therefore, we provide the rigorous proof of Proposition~\ref{pro:gamma0} in Appendix~\ref{apx:gamma0}. The proof of Proposition~\ref{pro:gamma0} is along the same line with the proof of Proposition~\ref{pro:gamma1}.

\section{Symmetric system}
The extant literature show that in a symmetric system, if all the service times are identical, introducing more flexibility would always be beneficial. However, here we have a new feature which is the prolonged service time at the non-dedicated server. Then people may wonder whether the full flexibility remains optimal with the prolonged service time at the non-dedicated server. In this section, we show that it depends on the value of the prolonged coefficient.

We first obtain the stationary probabilities for the \emph{symmetric full flexibility system} in the next proposition.
\begin{proposition}{\label{pro:simplified_model}}
When the system is symmetric, \ie $k = 1$, the stationary probabilities of the full flexibility system are
\begingroup
\allowdisplaybreaks
\begin{align*}
\pi^{k = 1}(0,0) = & \frac{\gamma^3+\gamma^2+2\gamma^2\rho}{\gamma^2(\rho+1)^3+\gamma^3(\rho+1)^2+\rho(\rho+\gamma)^2+2\gamma\rho^2(\rho+\gamma)}, \\
\pi^{k = 1}(0,1) = & \frac{\rho^2}{\gamma^2+\gamma+2\gamma\rho}\pi^{k = 1}(0,0), \\
 \pi^{k = 1}(0,2) = &\frac{\rho^2+(\gamma+1)\rho}{\gamma+2\rho+1}\pi^{k = 1}(0,0),\\
\pi^{k = 1}(1,0) = & \frac{\rho^2+(\gamma+1)\rho}{\gamma+2\rho+1}\pi^{k = 1}(0,0), \\
 \pi^{k = 1}(1,1) = & \frac{\rho^3+\gamma\rho^2}{\gamma^2+\gamma+2\gamma\rho}\pi^{k = 1}(0,0),\\
\pi^{k = 1}(1,2) = & \frac{\rho^3+(\gamma+1)\rho^2}{\gamma+2\rho+1}\pi^{k = 1}(0,0), \\
 \pi^{k = 1}(2,0) = & \frac{\rho^2}{\gamma^2+\gamma+2\gamma\rho}\pi^{k = 1}(0,0),\\
\pi^{k = 1}(2,1) = & \frac{\rho^3}{\gamma^3+\gamma^2+2\gamma^2\rho}\pi^{k = 1}(0,0), \\
 \pi^{k = 1}(2,2) = & \frac{\rho^3+\gamma\rho^2}{\gamma^2+\gamma+2\gamma\rho}\pi^{k = 1}(0,0).
\end{align*}
\endgroup
\end{proposition}
The result can be readily verified thus we omit the proof.


Next, we obtain the stationary probabilities for the \emph{symmetric partial flexibility system}. 

\begin{proposition}{\label{pro:probability_k=1}}
When the system is symmetric, \ie $k = 1$, the stationary probabilities of partial flexibility system are
\begin{align*}
&p^{k = 1}(0,0) \\
&= \frac{2\gamma^2\rho + 2\gamma^2 + 3\gamma\rho^2 + 6\gamma\rho + 2\gamma}{(\rho + 1)(2\gamma^2(\rho + 1)^2 + \gamma(2\rho^3 + 9 \rho^2 + 8\rho + 2) + 2\rho^2(\rho + 1))}, \\ 
&p^{k = 1}(1,0) = \frac{2\rho(\rho + 1)(\gamma + \rho + 1)}{2\gamma + 6\rho + 2\gamma\rho + 3\rho^2 + 2} p^{k = 1}(0,0), \\ &p^{k = 1}(0,1) =\frac{2\rho^2(\rho + 1)}{\gamma(2\gamma + 6\rho + 2\gamma\rho + 3\rho^2 + 2)} p^{k = 1}(0,0),\\
&p^{k = 1}(0,2) =\frac{2\rho(\gamma + 3\rho + \gamma\rho + \rho^2 + 1)}{2\gamma + 6\rho + 2\gamma\rho + 3\rho^2 + 2}p^{k = 1}(0,0),\\
&p^{k = 1}(1,1) =\frac{2\rho^2(\gamma + \rho)(\rho + 1)}{\gamma(2\gamma + 6\rho + 2\gamma\rho + 3\rho^2 + 2)} p^{k = 1}(0,0),\\
&p^{k = 1}(1,2) =\frac{\rho^2(2\gamma + 5\rho + 2\gamma\rho + 2\rho^2 + 2)}{2\gamma + 6\rho + 2\gamma\rho + 3\rho^2 + 2} p^{k = 1}(0,0).
 \end{align*}
\end{proposition}
The result can be readily verified thus we omit the proof.

When the system is symmetric, the throughputs can be obtained by substituting the corresponding stationary probabilities into Equations~\eqref{eq:tis}-\eqref{eq:tps},
\begin{align*}
&T^{k = 1}_{is} = \frac{2 \rho}{\rho + 1},  \\
&T^{k = 1}_{fs} = \\
&\frac{2\gamma\rho(2\gamma^2\rho + \gamma^2 + 2\gamma\rho^2 + 4\gamma\rho + \gamma + 2\rho^2)}{\gamma^3(\rho + 1)^2 + \gamma^2(\rho^3 + 5\rho^2 + 4\rho + 1) + 2\gamma \rho^2(\rho + 1) + \rho^3},\\
&T^{k = 1}_{ps} = \frac{2\rho(3\gamma^2\rho + 2\gamma^2 + 3\gamma\rho^2 + 7\gamma\rho + 2\gamma + \rho^2)}{2\gamma^2(\rho + 1)^2 + \gamma(2\rho^3 + 9\rho^2 + 8\rho + 2) + 2\rho^2(\rho + 1)}.
\end{align*}

Next, we investigate the ordering of the throughputs in the symmetric systems. 

\begin{proposition}{\label{pro:k1}}
When the system is symmetric, \ie $k = 1$, we have $T_{is}^{k = 1} < T_{ps}^{k = 1} < T_{fs}^{k = 1}$ for $\gamma \in (\frac{\rho}{\rho + 1},1)$ and $T_{fs}^{k = 1} < T_{ps}^{k = 1} < T_{is}^{k = 1}$ for $\gamma \in (0,\frac{\rho}{\rho + 1})$.
\end{proposition}

We see that in a symmetric system, whether the full flexibility remains optimal relies on the value of the prolonged coefficient. If the prolonged coefficient is close to $1$, \ie the negative consequences of redirecting the customers to the non-dedicated servers are not very serious, the full flexibility is always suggested. However, if the negative consequences of redirecting the customers to the non-dedicated servers are serious, \ie the prolonged coefficient is close to $0$, we see that working independently is the optimal choice. Another conclusion can also be drawn based on the proof of Proposition~\ref{pro:k1}.

\begin{proposition}{\label{pro:ordering_fix_gamma_kis1}}
When the system is symmetric,\ie $k = 1$, we have $T_{is}^{k = 1} < T_{ps}^{k = 1} < T_{fs}^{k = 1}$ for $\rho \in (0,\frac{\gamma}{1 - \gamma})$ and $T_{fs}^{k = 1} < T_{ps}^{k = 1} < T_{is}^{k = 1}$ for $\rho \in (\frac{\gamma}{1 - \gamma}, \infty)$.
\end{proposition}
Whether the full flexibility remains optimal in a symmetric system also depends on the workload in the system if the service time is prolonged at the non-dedicated server. Proposition~\ref{pro:ordering_fix_gamma_kis1} indicates that when the system is not very busy, more flexibility is welcome. However, when the system is relatively busy, it is better to keep them working independently.

This result verifies the observation~\cite{dijk2008pool} that in a heterogeneous system, more flexibility is not welcome when the traffic in the system is heavy. Therefore, when the effect of prolonged service time exists, even in a symmetric system, the decision makers should be very cautious when deciding how to introduce flexibility into the system and should not take it for granted that full flexibility is always optimal in a symmetric system.

The majority of the current literature considers the symmetric systems. On one hand, our results are consistent with the general conclusion of the current papers. Moreover, we extend the general conception to the asymmetric system with the prolonged service time at the non-dedicated server. However, it remains unclear that what the explicit ordering of the throughputs of the independent system, the full flexibility system and the partial flexibility system is when the system is allowed to be asymmetric and the service time at the non-dedicated server is prolonged. Moreover, when the service times are all identical, \ie $\gamma = 1$ and the system is symmetric, \ie $k = 1$, the choice for introducing flexibility becomes a ``0-1" decision, either full flexibility or no flexibility at all is preferred. Whether partial flexibility could also be optimal still remains unclear. The next section is devoted to providing comprehensive investigation of these problems.

\section{The general optimal flexibility design}{\label{sec:fix_rho}}
In total, there are $3$ system parameters which are allowed to change, the occupation rate $\rho$, the degree of asymmetry $k$ and the prolonged coefficient $\gamma$. We will focus on the triple of these $3$ parameters in the analysis later on. We are interested in determining the optimal flexibility scheme for $\rho \in (0,\infty)$, $k \in (0,1)$ and $\gamma \in (0,1)$. We first define the level sets $\mathcal{A}_r$, $\mathcal{A}_b$ and $\mathcal{A}_g$ to characterize the system parameters which make the throughputs of different systems identical.

\emph{The level sets}: For $\rho \in (0,\infty)$, $k \in (0,1)$, $\gamma \in (0,1)$, we call the set of $(\rho,k,\gamma)$ which makes the throughputs of the full flexibility system and the partial flexibility system identical to be $\mathcal{A}_r$ where $\mathcal{A}_r = \{(\rho,k,\gamma) | T_{fs}(\rho,k,\gamma) - T_{ps}(\rho,k,\gamma) = 0\}$. We call the set of $(\rho,k,\gamma)$ which makes the throughputs of the full flexibility system and the independent system identical to be $\mathcal{A}_b$ where $\mathcal{A}_b = \{(\rho,k,\gamma) | T_{fs}(\rho,k,\gamma) - T_{is}(\rho,k,\gamma) = 0\}$. We call the set of $(\rho,k,\gamma)$ which makes the throughputs of the partial flexibility system and the independent system identical to be $\mathcal{A}_g$ where $\mathcal{A}_g = \{(\rho,k,\gamma) | T_{ps}(\rho,k,\gamma) - T_{is}(\rho,k,\gamma) = 0\}$. Notice that the throughputs are the functions of $\rho$, $k$ and $\gamma$; the notations for the throughputs used above should be clear for the readers.

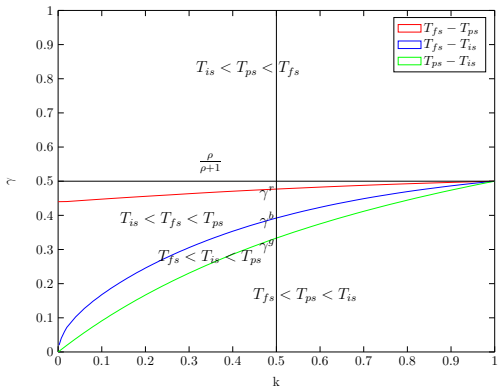
\begin{figure}[htbp]
\begin{center}
%
%
\definecolor{mycolor1}{rgb}{0.56250,1.00000,0.43750}%
\begin{tikzpicture}[scale = 0.5]

\begin{axis}[%
width=4.521in,
height=3.566in,
at={(0.758in,0.481in)},
scale only axis,
separate axis lines,
every outer x axis line/.append style={black},
every x tick label/.append style={font=\color{black}},
every x tick/.append style={black},
xmin=0,
xmax=1,
xlabel={k},
every outer y axis line/.append style={black},
every y tick label/.append style={font=\color{black}},
every y tick/.append style={black},
ymin=0,
ymax=1,
ylabel={$\gamma$},
axis background/.style={fill=white},
legend style={legend cell align=left, align=left, draw=black}
]

\addplot[area legend, draw=red]
table[row sep=crcr] {%
0	0.44\\
0.02	0.440413088774396\\
0.04	0.442218748655947\\
0.06	0.444003244144437\\
0.08	0.445763200757161\\
0.1	0.44749594404022\\
0.12	0.449199388826874\\
0.14	0.4508719455671\\
0.16	0.452512440982082\\
0.18	0.454120050780389\\
0.2	0.455694242560029\\
0.22	0.457234727332439\\
0.24	0.458741418358883\\
0.25687203089303	0.46\\
0.26	0.460240794657576\\
0.28	0.461853564077751\\
0.3	0.463422663381432\\
0.32	0.464949182758993\\
0.34	0.46643423804332\\
0.36	0.467878959069436\\
0.38	0.469284480132627\\
0.4	0.470651932211167\\
0.42	0.471982436672065\\
0.44	0.473277100222187\\
0.46	0.474537010904099\\
0.48	0.475763234966875\\
0.5	0.476956814468576\\
0.52	0.478118765488967\\
0.54	0.479250076850088\\
0.553455166165162	0.48\\
0.56	0.480384558808279\\
0.58	0.481555729008284\\
0.6	0.482693736420785\\
0.62	0.483799765537839\\
0.64	0.484874951122728\\
0.66	0.485920380247508\\
0.68	0.486937094294889\\
0.7	0.487926090914971\\
0.72	0.488888325929729\\
0.74	0.489824715179888\\
0.76	0.490736136310479\\
0.78	0.4916234304925\\
0.8	0.492487404079311\\
0.82	0.493328830197101\\
0.84	0.494148450269501\\
0.86	0.494946975477006\\
0.88	0.495725088152214\\
0.9	0.496483443112334\\
0.92	0.497222668930542\\
0.94	0.49794336914814\\
0.96	0.498646123429411\\
0.98	0.499331488661286\\
1	0.5\\
};
\addlegendentry{$T_{fs} - T_{ps}$}

\addplot[area legend, draw=blue]
table[row sep=crcr] {%
x	y\\
0.00220959078750193	0.02\\
0.00717044721923093	0.04\\
0.0146937521513777	0.06\\
0.02	0.0720460689251441\\
0.0252446309955065	0.08\\
0.0384053633631335	0.1\\
0.04	0.102483442167813\\
0.0542013288555959	0.12\\
0.06	0.127317641203655\\
0.0721136829695924	0.14\\
0.08	0.148663440804304\\
0.0920868276729245	0.16\\
0.1	0.16785942296742\\
0.114045224848972	0.18\\
0.12	0.185468466252514\\
0.137938460587578	0.2\\
0.14	0.201775609675907\\
0.16	0.217285516143393\\
0.163893877056741	0.22\\
0.18	0.232009088340471\\
0.191919697542543	0.24\\
0.2	0.245780946025259\\
0.22	0.258767657657702\\
0.2220586387457	0.26\\
0.24	0.271475131982071\\
0.254675046456633	0.28\\
0.26	0.283296725361488\\
0.28	0.294761023245569\\
0.289930562650011	0.3\\
0.3	0.305660070345426\\
0.32	0.316051525420499\\
0.32818276814765	0.32\\
0.34	0.326070605371063\\
0.36	0.335595329292138\\
0.36993355044787	0.34\\
0.38	0.344746603216248\\
0.4	0.353550099063413\\
0.41578334332447	0.36\\
0.42	0.361830025774605\\
0.44	0.370019541201462\\
0.46	0.37761713555988\\
0.466652601151923	0.38\\
0.48	0.385073796283316\\
0.5	0.392203442112922\\
0.52	0.398853981611566\\
0.523631461725693	0.4\\
0.54	0.405476364695841\\
0.56	0.41175795893702\\
0.58	0.417643253149756\\
0.588451582614216	0.42\\
0.6	0.423409774236887\\
0.62	0.428996536330164\\
0.64	0.434249848863035\\
0.66	0.439195840432628\\
0.663406403446374	0.44\\
0.68	0.444143097014163\\
0.7	0.448865533559557\\
0.72	0.45332451703647\\
0.74	0.45753916636605\\
0.752277429380412	0.46\\
0.76	0.461634895956064\\
0.78	0.465677128071933\\
0.8	0.469506804485989\\
0.82	0.473138322550428\\
0.84	0.476584895087624\\
0.86	0.479858663502277\\
0.860898242192139	0.48\\
0.88	0.483170353560186\\
0.9	0.486328235095421\\
0.92	0.489333012519589\\
0.94	0.492194146653951\\
0.96	0.49492038740923\\
0.98	0.497519835825862\\
1	0.5\\
};
\addlegendentry{$T_{fs} - T_{is}$}

\addplot[area legend, draw=green]
table[row sep=crcr] {%
x	y\\
0	0\\
0.02	0.0196264587238728\\
0.020416360273523	0.02\\
0.04	0.0385281451500507\\
0.0416986063826154	0.04\\
0.06	0.0567321369703541\\
0.063895494571881	0.06\\
0.08	0.0742587007510531\\
0.0870538349472438	0.08\\
0.1	0.0911229471953083\\
0.111217768138369	0.1\\
0.12	0.107336057855528\\
0.136428111519927	0.12\\
0.14	0.122906204214254\\
0.16	0.138005855921285\\
0.162843098350999	0.14\\
0.18	0.152727100257461\\
0.190584765025347	0.16\\
0.2	0.166838751264336\\
0.219522494392263	0.18\\
0.22	0.180340121034133\\
0.24	0.19371194051701\\
0.250108139068381	0.2\\
0.26	0.206508603416238\\
0.28	0.218792518488632\\
0.282090979292872	0.22\\
0.3	0.230946739315325\\
0.315860177715338	0.24\\
0.32	0.242499181606202\\
0.34	0.253880954193691\\
0.351454930734208	0.26\\
0.36	0.264829242501943\\
0.38	0.275483116866921\\
0.388992007640122	0.28\\
0.4	0.285850916618783\\
0.42	0.295885102421642\\
0.428672382114198	0.3\\
0.44	0.305686947618005\\
0.46	0.315189053618602\\
0.47068934100487	0.32\\
0.48	0.324433075154589\\
0.5	0.333474733919482\\
0.515224693456547	0.34\\
0.52	0.342164567042486\\
0.54	0.35080487585141\\
0.56	0.359004523592876\\
0.562543037741473	0.36\\
0.58	0.367229123616489\\
0.6	0.375114335684995\\
0.612991171257232	0.38\\
0.62	0.382786974548717\\
0.64	0.390393904327764\\
0.66	0.39765357438676\\
0.666750468457159	0.4\\
0.68	0.404869231605544\\
0.7	0.411907155118559\\
0.72	0.418642598733468\\
0.724198478979777	0.42\\
0.74	0.425400206920978\\
0.76	0.431957693470784\\
0.78	0.438249211425515\\
0.785787721153901	0.44\\
0.8	0.444542970995002\\
0.82	0.450690826617843\\
0.84	0.456602836025504\\
0.8519362721347	0.46\\
0.86	0.462424075713013\\
0.88	0.46821954755354\\
0.9	0.473804291165104\\
0.92	0.479188762469887\\
0.923121340784304	0.48\\
0.94	0.484632541890475\\
0.96	0.489932812477874\\
0.98	0.495052484354946\\
1	0.5\\
};
\addlegendentry{$T_{ps} - T_{is}$}

\addplot[area legend, draw=black]
table[row sep=crcr] {%
x	y\\
0	0.5\\
1   0.5\\
};

\addplot[area legend, draw=black]
table[row sep=crcr] {%
x	y\\
0.5	0\\
0.5   1\\
};

\node at (4cm,5cm) {\large{$\frac{\rho}{\rho + 1}$}};
\node at (5.5cm,4.2cm) {\large{$\gamma^r$}};
\node at (5.5cm,3.5cm) {\large{$\gamma^b$}};
\node at (5.5cm,2.8cm) {\large{$\gamma^g$}};


\node at (5cm,7.5cm) {\large{$T_{is} < T_{ps} < T_{fs}$}};
\node at (3cm,3.5cm) {\large{$T_{is} < T_{fs} < T_{ps}$}};
\node at (4cm,2.5cm) {\large{$T_{fs} < T_{is} < T_{ps}$}};
\node at (6.5cm,1.5cm) {\large{$T_{fs} < T_{ps} < T_{is}$}};

\end{axis}
\end{tikzpicture}%
\caption{An example of the level sets, comparison of the throughputs when $\rho = 1$ and an example of $\gamma^r$, $\gamma^b$ and $\gamma^g$ when $k = 0.5$.\label{fig:example}}
\end{center}
\end{figure}

We illustrate an example of the level sets when $\rho = 1$ while the other $2$ system parameters are allowed to change in Figure~\ref{fig:example}. In particular, the red, blue and green curves are the level sets $\mathcal{A}_r$, $\mathcal{A}_b$ and $\mathcal{A}_g$ respectively. Clearly, these level sets can be used to separate the parameter space in which the ordering of the throughputs could be determined. We see when $\rho = 1$, the level set $\mathcal{A}_r$ is a continuous curve (due to manipulation of the polynomials, see Appendix~\ref{apx:unique_intersection} for more details) which separates the parameter space into two parts. We may immediately wonder whether there would be more branches (from a geometric view) for the level set $\mathcal{A}_r$. Moreover, when the workload $\rho$ is also allowed to change, whether the properties we observed in Figure~\ref{fig:example} would remain. Therefore, we obtain the next lemma. For any $\rho \in (0,\infty)$, $k \in (0,1)$, we prove that there exists a unique $\gamma^r \in \mathcal{A}_r$, a unique $\gamma^b \in \mathcal{A}_b$ and a unique $\gamma^g \in \mathcal{A}_g$. An example of $\gamma^r$, $\gamma^b$ and $\gamma^g$ when $\rho = 1$ and $k = 0.5$ can be found in Figure~\ref{fig:example}. This result excludes the possibilities to have more algebraic branches for the red, green and blue curves. Therefore, we can conclude that each level set separates the parameter space into only $2$ parts, not more. Moreover, the ordering of the throughputs in each part is determined. This would be extremely helpful in deciding the ordering of the throughputs when the parameters change, based on which, the general optimal flexibility design can be determined consequently.

\begin{lemma}{\label{lem:unique_intersection}}
For any $\rho \in (0,\infty)$, $k \in (0,1)$, there exists a unique $\gamma^r \in \mathcal{A}_r$, a unique $\gamma^b \in \mathcal{A}_b$ and a unique $\gamma^g \in \mathcal{A}_g$. Moreover, we have $\gamma_g = \frac{k \rho}{k\rho + 1}$.
\end{lemma}

Moreover, we investigate the intersections of the level sets $\mathcal{A}_r$, $\mathcal{A}_b$ and $\mathcal{A}_g$. In fact, the intersection of any two sets must be the intersection of all three sets.

\begin{proposition}{\label{pro:intersections_k_gamma}}
If $(\rho,k,\gamma) \in \mathcal{A}_r \cap \mathcal{A}_b$, then we must have $(\rho,k,\gamma) \in \mathcal{A}_r \cap \mathcal{A}_b \cap \mathcal{A}_g$. Similarly, if $(\rho,k,\gamma) \in \mathcal{A}_r \cap \mathcal{A}_g$, then we must have $(\rho,k,\gamma) \in \mathcal{A}_r \cap \mathcal{A}_b \cap \mathcal{A}_g$. Moreover, if $(\rho,k,\gamma) \in \mathcal{A}_b \cap \mathcal{A}_g$, then we must have $(\rho,k,\gamma) \in \mathcal{A}_r \cap \mathcal{A}_b \cap \mathcal{A}_g$.
\end{proposition}

The proof of Proposition~\ref{pro:intersections_k_gamma} is intuitive thus we omit it here. For instance, the parameter triple $(\rho,k,\gamma)$ which makes $T_{fs} = T_{ps}$, $T_{fs} = T_{is}$ would guarantee that $T_{ps} = T_{is}$. In fact, we will show later that for $k \in (0,1)$, the intersection of the level sets is an empty set. This result excludes the possibility to separate the parameter space into more than $4$ parts, which leads to more complexity in deciding the ordering of the throughputs.

In the next lemma, we show that for $k \in [0,1]$, the intersection of all level sets can only occur when $k = 0$ or $k = 1$. We prove that when $k \in (0,1)$, the level sets $A^r$ and $A^g$ have no intersection. Applying Proposition~\ref{pro:intersections_k_gamma}, we know that there is no intersection for all $3$ level sets when $k \in (0,1)$. 

\begin{lemma}{\label{lem:kis0and1}}
When $k = 0$, we have $\gamma^r = \gamma^b = \gamma^g = 0$. Moreover, when $k = 1$, we have $\gamma^r = \gamma^b = \gamma^g = \frac{\rho}{\rho + 1}$. Moreover, when $k \in (0,1)$, we have $\gamma^r \neq \gamma^b$, $\gamma^r \neq \gamma^g$ and $\gamma^b \neq \gamma^g$.
\end{lemma}

In the next lemma, we show that the ordering of $\gamma^r$, $\gamma^b$ and $\gamma^g$ is fixed for any $\rho \in (0,\infty)$, $k \in (0,1)$ and $\gamma \in (0,1)$. Moreover, we prove that $\gamma^r$, $\gamma^b$ and $\gamma^g$ cannot exceed $\frac{\rho}{\rho +1}$.

\begin{lemma}{\label{lem:relation_gamma}}
When $k \in (0,1)$, we have $0 < \gamma^g < \gamma^b <\gamma^r <\frac{\rho}{\rho + 1}$.
\end{lemma}

This result implies that when $\gamma >\frac{\rho}{\rho + 1}$, the full flexibility is always the best. This is true because we know from Proposition~\ref{pro:gamma1} that when $\gamma = 1$, \ie all service times are identical, the full flexibility is always the best. The current result indicates that as long as the service time at the non-dedicated server does not prolong too much, the full flexibility would remain the optimal choice. Only if the service time at the non-dedicated server is prolonged over certain degree, the optimal flexibility scheme may vary.

Moreover, the result in Lemma~\ref{lem:relation_gamma} regulates the ordering of $\gamma^r$, $\gamma^b$ and $\gamma^g$ for any $\rho \in (0,\infty)$ and $k \in (0,1)$, which helps to determine the ordering of the throughputs when the parameters change.

\begin{theorem}[Ordering of the throughputs]{\label{thm:optimal}}
For any $\rho \in (0,\infty)$ and $k \in (0,1)$, the ordering of the throughputs is
\begin{enumerate}
\item We have $T_{fs} < T_{ps} < T_{is}$ for $\gamma \in (0, \frac{k \rho}{k \rho  + 1})$.
\item We have $T_{fs} < T_{is} < T_{ps}$ for $\gamma \in (\frac{k \rho}{k \rho  + 1}, \gamma^b)$.
\item We have $T_{is} < T_{fs} < T_{ps}$ for $\gamma \in (\gamma^b, \gamma^r)$.
\item We have $T_{is} < T_{ps} < T_{fs}$ for $\gamma \in (\gamma^r, 1)$.
\end{enumerate}
\end{theorem}

The illustration of the ordering of the throughputs when $\rho = 1$ can be found in Figure~\ref{fig:example}. Clearly, it follows immediately from Theorem~\ref{thm:optimal} that the optimal flexibility design can be determined when the parameters change.

\begin{corollary}[Optimal flexibility design]{\label{cor:optimal}}
For any $\rho \in (0,\infty)$ and $k \in (0,1)$,
\begin{enumerate}
\item Independent system is optimal for $\gamma \in (0, \frac{k \rho}{k \rho  + 1})$.
\item Partial flexibility system is optimal for $\gamma \in (\frac{k \rho}{k \rho +1}, \gamma^r)$.
\item Full flexibility system is optimal for $\gamma \in (\gamma^r, 1)$.
\end{enumerate}
\end{corollary}

This result is not only of mathematical importance, it also shows a general trend of the optimal flexibility scheme when the occupation rate, the degree of the asymmetry and the degree of the prolonged service time at the non-dedicated server change. When the practical system has the similar characteristics of the triple of the parameters we considered, the indicated optimal flexibility scheme has presented the optimal level of flexibility that may be introduced in real cases. Moreover, unlike the existing results which mainly focus on showing that the partial flexibility is almost as good as the full flexibility in a symmetric system, we provide evidence that the partial flexibility could outperform the full flexibility when the scenarios change, \eg the system is asymmetric and the service time is prolonged at the dedicated server.

\section*{Acknowledgement}
Yanting Chen acknowledges support through the NSFC grant $71701066$.

\bibliographystyle{elsarticle-num}
\bibliography{flexibility_bib}

\begin{thebibliography}{10}
\expandafter\ifx\csname url\endcsname\relax
  \def\url#1{\texttt{#1}}\fi
\expandafter\ifx\csname urlprefix\endcsname\relax\def\urlprefix{URL }\fi
\expandafter\ifx\csname href\endcsname\relax
  \def\href#1#2{#2} \def\path#1{#1}\fi

\bibitem{song2020capacity}
H.~Song, A.~Tucker, R.~Graue, S.~Moravick, J.~Yang, Capacity pooling in
  hospitals: The hidden consequences of off-service placement, Management
  Science 66~(9) (2020) 3825--3842.

\bibitem{jordan1995benefits}
W.~Jordan, S.~Graves, Principles on the benefits of manufacuring process
  flexibility, Management Science 41~(4) (1995) 577--594.

\bibitem{bassamboo2012flexibility}
A.~Bassamboo, R.~Randhawa, J.~Van~Mieghem, A little flexibility is all you
  need: On the asymptotic value of flexibility in parallel queuing systems with
  linear capacity sizing costs, Operations Research 60~(6) (2012) 1423--1435.

\bibitem{buzacott1996commonalities}
J.~Buzacott, Commonalities in reengineered business processes: Models and
  issues, Management Science 42 (1996) 768--782.

\bibitem{smith1981resource}
D.~Smith, W.~Whitt, Resource sharing for efficiency in traffic systems, Bell
  System Technical Journal 60~(1) (1981) 39--55.

\bibitem{chou2008process}
M.~Chou, C.~Teo, H.~Zheng, Process flexiblity: design, evaluation, and
  applications, Flexible Services and Manufacturing Journal 20~(1-2) (2008)
  59--94.

\bibitem{wang2019review}
S.~Wang, X.~Wang, J.~Zhang, A review of flexible processes and operations,
  Production and Operations Management (2019).
\newblock \href {https://doi.org/10.1111/poms.13101}
  {\path{doi:10.1111/poms.13101}}.

\bibitem{kleinrock1975queueing}
L.~Kleinrock, Queueing Systems, Vol. I: Theory, Wiley, New York, 1975.

\bibitem{wolff1982poisson}
R.~Wolff, Poisson arrivals see time averages, Operations Reserach 30 (1982)
  223--231.

\bibitem{dijk2008pool}
N.~van Dijk, E.~van~der Sluis, To pool or not to pool in call centers,
  Production and Operations Management 17~(3) (2008) 296--305.

\end{thebibliography}



%

\appendix
\renewcommand*{\thesection}{\Alph{section}}
\section{The balance equations for the full and partial flexibility systems}{\label{apx:balance_equations}}

The balance equations for the full flexibility system are
\begingroup
\allowdisplaybreaks
\begin{align*}
(\rho + k \rho) \pi(0,0) &= \pi(1,0) + \gamma \pi(0,1) + \pi(0,2) +\gamma \pi(2,0), \\
(\rho + k \rho + 1) \pi(1,0) &= \rho \pi(0,0) +\gamma \pi(1,1) + \pi(1,2), \\
(\rho + k \rho + \gamma) \pi(2,0) &= \gamma \pi(2,1) + \pi(2,2), \\
(\rho + k \rho  + \gamma) \pi(0,1) &= \pi(1,1) + \gamma \pi(2,1), \\
(1 +\gamma) \pi(1,1) &= \rho \pi(1,0) +  \rho \pi(0,1), \\
2 \gamma \pi(2,1) &= \rho \pi(2,0) + k \rho \pi(0,1), \\
(\rho + k \rho +1) \pi(0,2) &= k \rho \pi(0,0) + \pi(1,2) + \gamma \pi(2,2), \\
2 \pi(1,2) &= \rho \pi(0,2) + k \rho \pi(1,0), \\
(\gamma + 1) \pi(2,2) &= k \rho \pi(2,0) + k \rho \pi(0,2).
\end{align*}
\endgroup

The balance equations for the partial flexibility system are
\begin{align*}
( \rho + k \rho) p(0,0) &= p(1,0) + \gamma p(0,1) + p(0,2),\\
(\rho + k \rho + 1) p(1,0) &= \rho p(0,0) + \gamma p(1,1) + p(1,2), \\
2 p(1,2) &= \rho p(0,2) + k \rho p(1,0), \\
(\rho + 1) p(0,2) &= k \rho p(0,0) + p(1,2),\\
(1 +\gamma) p(1,1) &= \rho (p(0,1) +p(1,0)),\\
( \rho + \gamma) p(0,1) &= p(1,1).
\end{align*}

\section{Proof of Proposition~\ref{pro:gamma1}}
\begin{proof}
We denote the nominator and the denominator of $T^{\gamma = 1}_{fs}$ by $T^{\gamma = 1,n}_{fs}$ and $T^{\gamma = 1,d}_{fs}$, the nominator and the denominator of $T^{\gamma = 1}_{ps}$ by $T^{\gamma = 1,n}_{ps}$ and $T^{\gamma = 1,d}_{ps}$, the nominator and the denominator of $T^{\gamma = 1}_{is}$ by $T^{\gamma = 1,n}_{is}$ and $T^{\gamma = 1,d}_{is}$. When $\gamma= 1$, we have 
\begin{align*}
&T_{fs}^{\gamma = 1,n} T_{ps}^{\gamma = 1,d} - T_{ps}^{\gamma = 1,n} T_{fs}^{\gamma = 1,d}\\
=& k\rho^2(k^3\rho^2 + 3k^2\rho^2 + 4k^2\rho + 3k\rho^2 + 6k\rho + 4k + \rho^2 + 2\rho).\\
&T_{fs}^{\gamma = 1,n} T_{is}^{\gamma = 1,d} - T_{is}^{\gamma = 1,n} T_{fs}^{\gamma = 1,d} = \rho^2(k^2 + 1)(\rho + k\rho + 2).\\
&T_{ps}^{\gamma = 1,n} T_{is}^{\gamma = 1,d} - T_{is}^{\gamma = 1,n} T_{ps}^{\gamma = 1,d} = \rho^2(\rho + 1)(\rho + k\rho + 2).
\end{align*}
We have $T_{fs}^{\gamma = 1,d} > 0$, $T_{ps}^{\gamma = 1,d} > 0$ and $T_{is}^{\gamma = 1,d} > 0$ due to $\rho \in (0,\infty)$ and $k \in [0,1]$, therefore we have $T^{\gamma = 1}_{is} < T^{\gamma = 1}_{ps} < T^{\gamma = 1}_{fs}$ when $\gamma = 1$, which completes the proof.
%
%
\end{proof}

\section{Proof of Proposition~\ref{pro:gamma0}}{\label{apx:gamma0}}
\begin{proof}
When $\gamma = 0$, the stationary probabilities of the full flexibility system are
\begin{align*}
 &\pi^{\gamma = 0}(0,0) = 0, \quad \pi^{\gamma = 0}(1,0) = 0, \quad \pi^{\gamma = 0}(0,1) = 0,\\
 &\pi^{\gamma = 0}(0,2) = 0, \quad \pi^{\gamma = 0}(2,0) = 0, \quad \pi^{\gamma = 0}(1,1) = 0,\\
 &\pi^{\gamma = 0}(1,2) = 0, \quad \pi^{\gamma = 0}(2,1) = 1, \quad \pi^{\gamma = 0}(2,2) = 0.
\end{align*}

When $\gamma= 0$, the stationary probabilities of the partial flexibility system are
\begin{align*}
p^{\gamma = 0}(0,0) = 0, \quad p^{\gamma = 0}(1,0) = 0, \quad p^{\gamma = 0}(0,1) = \frac{1}{\rho + 1},\\
p^{\gamma = 0}(0,2) = 0, \quad p^{\gamma = 0}(1,1) = \frac{\rho}{\rho + 1}, \quad p^{\gamma = 0}(1,2) = 0.
\end{align*}
When $\gamma = 0$, the throughputs of the $3$ flexible systems are: $T_{fs}^{\gamma = 0} = 0$, $T_{ps}^{\gamma = 0} = \frac{\rho}{\rho + 1}$ and $T_{is}^{\gamma = 0} = \frac{\rho}{\rho + 1} + \frac{k \rho}{k \rho + 1}$. Therefore, we conclude that when $\gamma = 0$, we have $T^{\gamma = 0}_{fs} < T^{\gamma = 0}_{ps} \leq T^{\gamma = 0}_{is}$, the equality sign holds when $k = 0$.
\end{proof}

\section{Proof of Proposition~\ref{pro:k1}}
\begin{proof}
We denote the nominator and denominator of $T_{fs}^{k = 1}$ by $T_{fs}^{k = 1,n}$ and $T_{fs}^{k = 1,d}$, the nominator and denominator of $T_{ps}^{k = 1}$ by $T_{ps}^{k = 1,n}$ and $T_{ps}^{k = 1,d}$, the nominator and denominator of $T_{is}^{k = 1}$ by $T_{is}^{k = 1,n}$ and $T_{is}^{k = 1,d}$. We now investigate the equation $T_{fs}^{k = 1} - T_{ps}^{k = 1} = 0$. We have 
\begin{align*}
&T_{fs}^{k = 1, n} T_{ps}^{k = 1,d} - T_{ps}^{k = 1,n} T_{fs}^{k = 1, d}\\
=&2\rho^2(\rho + 1)(\gamma - \frac{\rho}{\rho + 1})(\gamma^4 + \rho\gamma^4 + 5\rho\gamma^3 + 2\rho^2\gamma^3 + 2\gamma^3 + \\ &6\rho\gamma^2 + 6\rho^2\gamma^2 + \gamma^2 + \rho^3\gamma^2 + 6\rho^2\gamma + 2\rho^3\gamma + 2\rho\gamma + \rho^3).
\end{align*} 
The assumptions $\rho \in (0,\infty)$ and $\gamma \in [0,1]$ imply that $T_{fs}^{k = 1,d} > 0$ and $T_{ps}^{k = 1,d} > 0$. Therefore, we obtain that $T_{fs}^{k = 1} > T_{ps}^{k = 1}$ for $\gamma \in (\frac{\rho}{\rho + 1},1)$ and $T_{fs}^{k = 1} < T_{ps}^{k = 1}$ for $\gamma \in (0,\frac{\rho}{\rho + 1})$. 

We then investigate the equation $T_{fs}^{k = 1} - T_{is}^{k = 1} = 0$, we have
\begin{align*}
&T_{fs}^{k = 1,n} T_{is}^{k = 1,d} -  T_{is}^{k = 1,n} T_{fs}^{k = 1,d} \\
= &2\rho^2 (\rho + 1)(\gamma - \frac{\rho}{\rho + 1})(\gamma + \rho)(\gamma + 1).
\end{align*}
The assumptions $\rho \in (0,\infty)$ and $\gamma\in [0,1]$ imply that $T_{is}^{k = 1,d} >0$ as well. Therefore, we obtain that $T_{fs}^{k = 1} > T_{is}^{k = 1}$ for $\gamma \in (\frac{\rho}{\rho + 1},1)$ and $T_{fs}^{k = 1} < T_{is}^{k = 1}$ for $\gamma \in (0,\frac{\rho}{\rho + 1})$. 

Finally, we investigate the equation $T_{ps}^{k = 1} - T_{is}^{k = 1} = 0$, we have
\begin{align*}
&T_{ps}^{k = 1,n} T_{is}^{k = 1,d} -  T_{is}^{k = 1,n} T_{ps}^{k = 1,d} \\
=  &2\rho^2 (\rho + 1)(\gamma - \frac{\rho}{\rho + 1})(\gamma + \rho + 1).
\end{align*}
We know that $T_{ps}^{k = 1,d} >0$ and $T_{is}^{k = 1,d} >0$ for $\rho \in (0,\infty)$ and $\gamma \in [0,1]$. Therefore, we obtain that $T_{ps} > T_{is}$ for $\gamma \in (\frac{\rho}{\rho + 1},1)$ and $T_{ps} < T_{is}$ for $\gamma \in (0,\frac{\rho}{\rho + 1})$, which completes the proof.


\end{proof}

\section{Proof of Lemma~\ref{lem:unique_intersection}}{\label{apx:unique_intersection}}
\begin{proof}
By solving the balance equations for the full and partial flexibility systems together with the normalization requirement, we obtain the stationary probabilities of the general full and partial flexibility systems. We are able to obtain the explicit expressions for $T_{fs}$ and $T_{ps}$ by inserting the corresponding stationary probabilities into Equations~\eqref{eq:tis}-\eqref{eq:tps}. We define $r(\rho,k,\gamma) = T_{fs}(\rho,k,\gamma) - T_{ps}(\rho,k,\gamma)$, $b(\rho,k,\gamma) = T_{fs}(\rho,k,\gamma) - T_{is}(\rho,k,\gamma)$ and $g(\rho,k,\gamma) = T_{ps}(\rho,k,\gamma) - T_{is}(\rho,k,\gamma)$. We denote the nominator and the denominator of $T_{fs}$ by $T^{n}_{fs}$ and $T^{d}_{fs}$, the nominator and the denominator of $T_{ps}$ by $T^{n}_{ps}$ and $T^{d}_{ps}$. We define $R(\rho, k, \gamma) = R_1(\rho, k, \gamma) - R_2(\rho, k, \gamma)$ where $R_1(\rho, k, \gamma) = T^{n}_{fs} T^{d}_{ps}$ and $R_2(\rho, k,\gamma) = T^{n}_{ps} T^{d}_{fs}$. From Proposition~\ref{pro:gamma1} we know that when $\gamma = 1$, we have $T^{\gamma = 1}_{fs} - T^{\gamma = 1}_{ps} >0$, which means $R(\rho,k,1) >0$ because $T^{\gamma = 1,d}_{fs} >0$ and $T^{\gamma = 1,d}_{ps} >0$ for $\rho \in (0,\infty)$ and $k \in (0,1)$. From Proposition~\ref{pro:gamma0} we know that when $\gamma = 0$, we have $T^{\gamma = 0}_{fs} - T^{\gamma = 0}_{ps} <0$, which means $R(\rho,k,0) <0$ because $T^{\gamma = 0,d}_{fs} >0$ and $T^{\gamma = 0,d}_{ps} >0$ for $\rho \in (0,\infty)$ and $k \in (0,1)$. Moreover, for any fixed $\rho \in (0,\infty)$ and $k \in (0,1)$, we know that $R(\rho, k, \gamma)$ is a polynomial of $\gamma$, hence $R(\rho,k,\gamma)$ is continuous in $\gamma$. For $\rho \in (0,\infty)$, $k \in (0,1)$ and $\gamma \in (0,1)$, we have $\frac{\partial R_{1}(\rho, k,\gamma)}{\partial \gamma} >0$, $\frac{\partial R_{2}(\rho, k, \gamma)}{\partial \gamma} >0$, 
$\frac{\partial^2 R_{1}(\rho, k, \gamma)}{\partial \gamma^2} >0$ and $\frac{\partial^2 R_{2}(\rho, k, \gamma)}{\partial \gamma^2} >0$. The verification is straightforward and cumbersome, hence we omit it here. Therefore, we conclude that $R_{1}(\rho, k, \gamma)$ and $R_{2}(\rho, k, \gamma)$ are monotonically increasing and convex in $\gamma$ for $\gamma \in (0,1)$. As a result, we assert that for any $\rho \in (0,\infty)$ and $k \in (0,1)$, there is a unique root for $\gamma$ in $(0,1)$ to have $T_{fs} - T_{ps} = 0$. Then we conclude that for any $\rho \in (0,\infty)$, $k \in (0,1)$, there exists a unique $\gamma^r \in \mathcal{A}_r$.

Similarly, we denote the nominator and the denominator of $T_{is}$ by $T^{n}_{is}$ and $T^{d}_{is}$, we define $B(\rho, k, \gamma) = B_1(\rho, k, \gamma) - B_2(\rho, k, \gamma)$ where $B_1(\rho, k, \gamma) = T^{n}_{fs} T^{d}_{is}$ and $B_2(\rho, k,\gamma) = T^{n}_{is} T^{d}_{fs}$. We know from Proposition~\ref{pro:gamma1} that when $\gamma = 1$, we have $T^{\gamma = 1}_{fs} - T^{\gamma = 1}_{is} >0$, which means $B(\rho,k,1) > 0$ because $T^{\gamma = 1,d}_{fs} >0$ and $T^{\gamma = 1,d}_{is} >0$ for $\rho \in (0,\infty)$ and $k \in (0,1)$. From Proposition~\ref{pro:gamma0} we know that when $\gamma = 0$, we have $T^{\gamma = 0}_{fs} - T^{\gamma = 0}_{is} <0$, which means $B(\rho,k,0) <0$ because $T^{\gamma = 0,d}_{fs} >0$ and $T^{\gamma = 0,d}_{is} >0$ for $\rho \in (0,\infty)$ and $k \in (0,1)$. Moreover, for any fixed $\rho \in (0,\infty)$ and $k \in (0,1)$, we know that $B(\rho, k, \gamma)$ is a polynomial of $\gamma$, hence $B(\rho,k,\gamma)$ is continuous in $\gamma$. For $\rho \in (0,\infty)$, $k \in (0,1)$ and $\gamma \in (0,1)$, we have $\frac{\partial B_{1}(\rho, k,\gamma)}{\partial \gamma} >0$, $\frac{\partial B_{2}(\rho, k, \gamma)}{\partial \gamma} >0$, $\frac{\partial^2 B_{1}(\rho, k, \gamma)}{\partial \gamma^2} >0$ and $\quad \frac{\partial^2 B_{2}(\rho, k, \gamma)}{\partial \gamma^2} >0$.  The verification is straightforward and cumbersome, hence we omit it here. Therefore, we conclude that $B_{1}(\rho, k, \gamma)$ and $B_{2}(\rho, k, \gamma)$ are monotonically increasing and convex in $\gamma$ for $\gamma \in (0,1)$. As a result, we assert that for any $\rho \in (0,1)$ and $k \in (0,1)$, there is a unique root for $\gamma$ in $(0,1)$ to have $T_{fs} - T_{is} = 0$. Then we conclude that for any $\rho \in (0,\infty)$, $k \in (0,1)$, there exists a unique $\gamma^b \in \mathcal{A}_b$.

Finally, we define $G(\rho, k, \gamma) = G_1(\rho, k, \gamma) - G_2(\rho, k, \gamma)$ where $G_1(\rho, k, \gamma) = T^{n}_{ps} T^{d}_{is}$ and $G_2(\rho, k,\gamma) = T^{n}_{is} T^{d}_{ps}$ . It can be readily verified that
\begin{align*}
G(\rho, k,\gamma) &= T_{ps}^{n} T_{is}^{d} - T_{is}^{n} T_{ps}^d \\
         &=\rho^2(\rho + 1)(\rho + k\rho + 2)(\gamma + \rho + 1)(\gamma - k\rho + \gamma k\rho).
\end{align*}
For $\rho \in (0,\infty)$, $k \in (0,1)$ and $\gamma \in (0,1)$, we have $\rho^2(\rho + 1)(\rho + k\rho + 2)(\gamma + \rho + 1) > 0$, $T^{d}_{ps} >0$ and $T^{d}_{is} >0$, which can be readily verified. Therefore, the set $\mathcal{A}_g$ is determined by solving $\gamma - k\rho + \gamma k\rho = 0$, which is $\gamma = \frac{k \rho}{k\rho + 1}$.
Therefore, we have $\mathcal{A}_g = \{(\rho,k,\gamma)| \gamma = \frac{k \rho}{k\rho + 1}\}$, \ie $\gamma^g = \frac{k \rho}{k \rho +1}$.
\end{proof}
%
%
%
\section{Proof of Lemma~\ref{lem:kis0and1}}
\begin{proof}
When $k = 0$, it can be verified that the throughputs of the $3$ systems are 
\begingroup
\allowdisplaybreaks
\begin{align*}
&T^{k = 0}_{is} =  \frac{\rho}{\rho + 1}, T^{k = 0}_{fs} =  \frac{\rho(\gamma^2\rho + \gamma^2 + \gamma\rho^2 + 3\gamma\rho + \gamma + \rho^2)}{(\rho + 1)(\gamma^2 + 2\gamma\rho + \gamma + \rho^2)},\\
&T^{k = 0}_{ps} =\frac{\rho(\gamma^2\rho + \gamma^2 + \gamma\rho^2 + 3\gamma\rho + \gamma + \rho^2)}{(\rho + 1)(\gamma^2 + 2\gamma\rho + \gamma + \rho^2)}.
\end{align*}
\endgroup
Because the expressions for $T_{fs}^{k = 0}$ and $T_{ps}^{k = 0}$ are identical, we conclude that when $k = 0$ we have $T_{fs}^{k = 0} = T_{ps}^{k = 0}$. Therefore, we have $\gamma^r = 0$ when $k = 0$.

We denote the nominator and the denominator of $T_{is}^{k = 0}$ by $T_{is}^{k = 0,n}$ and $T_{is}^{k = 0,d}$, the nominator and the denominator of $T_{fs}^{k = 0}$ by $T_{fs}^{k = 0,n}$ and $T_{fs}^{k = 0,d}$, we have
\begin{align*}
T_{fs}^{k = 0,n} T_{is}^{k = 0,d} - T_{is}^{k = 0,n} T_{fs}^{k = 0,d}=\rho^2(\rho + 1)\gamma(\gamma + \rho + 1).
\end{align*}
The assumption $\rho \in (0,\infty)$ implies that $\gamma^b = 0$ when $k = 0$. Moreover, because $T_{fs}^{k = 0}$ and $T_{ps}^{k = 0}$ are identical when $k = 0$, we have $\gamma^g = 0$ as well when $k = 0$. When $k = 1$, it follows immediately from Proposition~\ref{pro:k1} that $\gamma_r = \gamma_b = \gamma_g = \frac{\rho}{\rho + 1}$.

We now prove that when $k \in (0,1)$, the intersection of the level sets $A_{r}$ and $A_{g}$ is an empty set. We know that $\gamma^g = \frac{k \rho}{k \rho + 1}$, substituting $\gamma = \frac{k \rho}{k \rho + 1}$ in to the polynomial $R(\rho,k,\gamma)$ we have $R(\rho,k,\frac{k \rho}{k \rho + 1}) = (k - 1)R'(\rho,k,\frac{k \rho}{k \rho + 1})$ where $R'(\rho,k,\frac{k \rho}{k \rho + 1})$ is a non-negative expression, which can be readily verified. Therefore, we have $R(\rho,k,\frac{k \rho}{k \rho + 1}) < 0$ for $\rho \in (0,\infty)$, $k \in (0,1)$ and $\gamma \in (0,1)$. Moreover, it can be readily verified that $T_{fs}^{d} > 0$ and $T_{ps}^d >0$ for $\rho \in (0,\infty)$, $k \in (0,1)$ and $\gamma \in (0,1)$. Therefore, we know that the intersection of the level sets $A_{r}$ and $A_{g}$ is an empty set for $\rho \in (0,\infty)$, $k \in (0,1)$ and $\gamma \in (0,1)$. Applying Proposition~\ref{pro:intersections_k_gamma}, we conclude that when $k \in (0,1)$, we have $\gamma^r \neq \gamma^g$, $\gamma^r \neq \gamma^b$ and $\gamma^g \neq \gamma^b$.

\end{proof}
\section{Proof of Lemma~\ref{lem:relation_gamma}}
\begin{proof}
For any $\rho \in (0,\infty)$ and $\gamma \in (0,1)$, when $k \in (0,1)$, we now investigate the ordering of $\gamma^{r}$, $\gamma^{b}$ and $\gamma^{g}$. We first show that it is not possible to have $\gamma^b < \gamma^g < \gamma^r$, $\gamma^b < \gamma^r < \gamma^g$, $\gamma^g < \gamma^r < \gamma^b$ and $\gamma^r < \gamma^g < \gamma^b$ by drawing contradictions. 

If $\gamma^b < \gamma^g < \gamma^r$, we consider the triple $(\rho, k, \gamma^{b})$, we know that $r(\rho, k, \gamma^{b}) = T_{fs}(\rho, k, \gamma^{b}) - T_{ps}(\rho, k, \gamma^{b}) < 0$ and $g(\rho, k, \gamma^{b}) = T_{ps}(\rho, k, \gamma^{b}) - T_{is}(\rho, k, \gamma^{b}) < 0$ due to Proposition~\ref{pro:gamma1} and Proposition~\ref{pro:gamma0}. This is true because for any fixed $\rho$, we know that each level set has separated the parameter space $k \in (0,1)$ and $\gamma \in (0,1)$ into two parts. Deploying Proposition~\ref{pro:gamma1} and Proposition~\ref{pro:gamma0}, the ordering of any two throughputs of the different flexibility systems could be consequently determined in each part resulted by the separation.  Therefore, we have
\begingroup
\allowdisplaybreaks
\begin{align*} 
b(\rho, k, \gamma^{b}) = & T_{fs}(\rho, k, \gamma^{b}) - T_{is}(\rho, k, \gamma^{b})\\
                     = & (T_{fs}(\rho, k, \gamma^{b}) - T_{ps}(\rho, k, \gamma^{b})) + \\
                       &(T_{ps}(\rho, k, \gamma^{b}) - T_{is}(\rho, k, \gamma^{b}))\\
                     <&0,
\end{align*}
\endgroup
which contradicts to the fact that $b(\rho, k, \gamma^{b}) = 0$. The contradiction for the case $\gamma^b < \gamma^r < \gamma^g$ can be drawn similarly.

If $\gamma^g < \gamma^r < \gamma^b$, we consider the triple $(\rho, k, \gamma^{b})$, we know that $r(\rho, k, \gamma^{b}) = T_{fs}(\rho, k, \gamma^{b}) - T_{ps}(\rho, k, \gamma^{b}) > 0$ and $g(\rho, k, \gamma^{b}) = T_{ps}(\rho, k, \gamma^{b}) - T_{is}(\rho, k, \gamma^{b}) > 0$ due to Proposition~\ref{pro:gamma1} and Proposition~\ref{pro:gamma0}. Therefore, we have
\begingroup
\allowdisplaybreaks
\begin{align*} 
b(\rho, k, \gamma^{b}) = & T_{fs}(\rho, k, \gamma^{b}) - T_{is}(\rho, k, \gamma^{b})\\
                     = & (T_{fs}(\rho, k, \gamma^{b}) - T_{ps}(\rho, k, \gamma^{b})) + \\
                       & (T_{ps}(\rho, k, \gamma^{b}) - T_{is}(\rho, k, \gamma^{b}))\\
                     >&0,
\end{align*}
\endgroup
which contradicts to the fact that $b(\rho, k, \gamma^{b}) = 0$. The contradiction for the case $\gamma^r < \gamma^g < \gamma^b$ can be drawn similarly.
 
Therefore, for $k \in (0,1)$ we can have either $\gamma^r < \gamma^b < \gamma^g$ or $\gamma^g < \gamma^b < \gamma^r$. Next, we show that when $k \in (0,1)$, we can only have $\gamma^g < \gamma^b < \gamma^r$. 
From the proof of Lemma~\ref{lem:unique_intersection}, it can be readily verified that $r(\rho,k,\frac{k \rho}{1 + k \rho}) < 0$ when $k \in (0,1)$, again using Proposition~\ref{pro:gamma1} and Proposition~\ref{pro:gamma0}, we conclude that for any $\rho \in (0,\infty)$ when $k \in (0,1)$, we have $\gamma^g < \gamma^b < \gamma^r$.

Inserting $\gamma = \frac{\rho}{\rho + 1}$ into $R(\rho,k,\gamma)$, we have
\begin{align*}
R(\rho,k,\frac{\rho}{\rho + 1}) = (1-k)\bar{R}(\rho, k, \frac{\rho}{\rho + 1}),
\end{align*}
where $\bar{R}(\rho, k, \frac{\rho}{\rho + 1})$ is a non-negative expression, which can be readily verified. Moreover, we know that $T^{d}_{fs} >0$ and $T^{d}_{ps} >0$ for $\rho \in (0,\infty)$, $k \in (0,1)$ and $\gamma \in (0,1)$, then we have $r(\rho,k,\frac{\rho}{\rho + 1}) > 0$ for $k \in (0,1)$. Again applying Proposition~\ref{pro:gamma1} and Proposition~\ref{pro:gamma0}, we conclude that when $k \in (0,1)$, we have $0 < \gamma^g < \gamma^b < \gamma^r < \frac{\rho}{\rho + 1}$, which completes the proof.

\end{proof}
%
%
\section{Proof of Theorem~\ref{thm:optimal}}
\begin{proof}
From Proposition~\ref{pro:gamma0} we know that for $\rho \in (0,\infty)$ and $k \in (0,1)$, when $\gamma = 0$, we have $T^{\gamma = 0}_{fs} < T^{\gamma = 0}_{ps} < T^{\gamma = 0}_{is}$. Therefore, for any $\rho \in (0,\infty)$, $k \in (0,1)$ and $\gamma \in (0,1)$, we have $T_{fs} < T_{ps}$ when $\gamma < \gamma^r$, $T_{fs} < T_{is}$ when $\gamma <\gamma^b$ and $T_{ps} <T_{is}$ when $\gamma < \gamma^g$. Proposition~\ref{pro:gamma1} indicates that for $\rho \in (0,\infty)$ and $k \in (0,1)$, when $\gamma = 1$, we have $T^{\gamma = 1}_{is} < T^{\gamma = 1}_{ps} < T^{\gamma = 1}_{fs}$. Therefore, for any $\rho \in (0,\infty)$, $k \in (0,1)$ and $\gamma \in (0,1)$, we have $T_{fs} >T_{ps}$ when $\gamma > \gamma^r$, $T_{fs} > T_{is}$ when $\gamma > \gamma^b$ and $T_{ps} > T_{is}$ when $\gamma > \gamma^g$. The summarization of the inequalities obtained above would complete the proof.
\end{proof}

\end{document}